\documentstyle[epsfig]{lamuphys}
\makeatletter
\let\chapter\hid@chapter
\makeatother
\begin{document}
\pagenumbering{arabic}
\title{Low Energy Experiments on $\pi$-$\pi$ Scattering}

\author{Dinko\,Po\v{c}ani\'c}
\institute{Department of Physics, University of Virginia, Charlottesville,
VA 22901, USA}

\maketitle

\begin{abstract}

General interest in a precise determination of the threshold $\pi$-$\pi$
amplitudes has recently increased markedly due to a controversy regarding
the size of $\langle 0|\bar{q}q|0\rangle$, the scalar quark condensate.
This paper examines the current experimental information on the $\pi$-$\pi$
scattering lengths, in particular the recent low energy $\pi N\to\pi\pi N$
data from several laboratories and the related application of the
Chew--Low--Goebel technique well below 1\,GeV/c momentum.  It appears that
uncertainties related to the treatment of non-pion-exchange backgrounds in
these studies do not yet allow an unambiguous resolution of the
$\langle 0|\bar{q}q|0\rangle$ size.  However, near-term prospects for new
model-independent results of improved precision are very good.

\end{abstract}

\section{Motivation \label{sec:motiv}}

Pion-pion scattering at threshold is uniquely sensitive to the explicit
chiral symmetry breaking (ChSB) portion of the strong interaction and has,
for this reason, been the subject of detailed study for over thirty years,
both theoretically and experimentally.  After QCD gained universal
acceptance as the theory of the strong interaction, long-time controversies
regarding the mechanism of the explicit breaking of chiral symmetry were
laid to rest and the Weinberg picture [\cite{wein:1}] was recognized as
valid at the tree level, providing a firm prediction for $a(\pi\pi)$, the
pion-pion scattering lengths.
 
However, QCD is not directly applicable at low energies, except numerically
on the lattice, which has not yet been established as a practical and
reliable calculational method.  Thus, knowing $a(\pi\pi)$, the $\pi$-$\pi$
scattering lengths, precisely has remained an important goal, as these
quantities provide a direct and sensitive constraint on parameters of the
available effective low energy lagrangians.  This is of particular
importance for the chiral perturbation theory (ChPT) approach which
provides a systematic framework for the treatment of low energy strong
interactions in terms of diagrams with increasing powers of momentum and
mass [\cite{gas:leu:1}].  Consequently, improved calculations including
one-loop [\cite{gas:leu:2}] and two-loop [\cite{bij:col:1}] diagrams have
been performed using standard ChPT.

Recently, however, a less restrictive version of ChPT was formulated by the
Orsay group, referred to as the generalized chiral perturbation theory
(GChPT) [\cite{ste:etal:1}].  This approach makes fewer theoretical
assumptions and consequently has more parameters than the standard ChPT for
the lagrangian terms of a given power of momentum or mass.  As in standard
ChPT, all parameters need to be constrained by data.  A particularly
interesting possibility that is allowed in GChPT concerns the very
mechanism of chiral symmetry breaking, as follows.

The standard picture of ChSB assumes a strong scalar quark condensate:
%$\langle 0 |\bar{q}q |0\rangle$:
\begin{equation}
     -\langle 0 |\bar{q}q |0\rangle \gg F_\pi^3 \enspace, \label{eq:strong}
\end{equation}
where $F_\pi \simeq 92$\,MeV is the pion decay constant.  In the standard
ChPT calculation, which relies on the above assumption, the s-wave $\pi\pi$
scattering lengths $a_{l=0}^I(\pi\pi)$ are predicted to be (including terms
with up to two loops [\cite{bij:col:1}]): 
\begin{equation}
      a_0^0 \simeq 0.21\ \mu^{-1} \qquad {\rm and} \qquad
      a_0^2 \simeq -0.041\ \mu^{-1} \enspace ,  \label{eq:chpt:apipi}
\end{equation}
where $I=0,2$ are the allowed values of dipion isospin and $\mu$ is the
charge-indepen\-dent pion mass. 

The Orsay group has argued for some time that the assumption in
(\ref{eq:strong}) is not clearly justified by the available experimental
evidence, and has claimed that a much weaker scalar quark condensate must,
in principle, be allowed [\cite{ste:etal:2},\cite{ste:etal:1}].  The
consequences of such a scenario are many, not the least of which are
radically different light quark mass ratios than the ones generally
accepted now [\cite{ste:etal:2},\cite{kne:mou:1}].  The only practical
observables sensitive to the size of $\langle 0|\bar{q}q|0\rangle$ are the 
s-wave $\pi\pi$ scattering lengths.

In particular, using the GChPT formalism and a weak scalar quark
condensate, the Orsay group found that the most likely value of
$a_0^0(\pi\pi)$ (calculated including one and two loop diagrams) would be
$\sim$0.27\,$\mu^{-1}$ [\cite{kne:mou:1},\cite{kne:mou:2}], about 30\,\%
higher than the standard ChPT calculation.  Clearly, a measurement of the
s-wave $\pi\pi$ scattering length with about 10\,\% precision is required
in order to differentiate experimentally between the two theoretical
results.

Although there have been many attempts at evaluating the $\pi\pi$
scattering lengths from available data over the years, the result generally 
accepted as most reliable is based on a comprehensive phase shift analysis
of peripheral $\pi N\to\pi\pi N$ reactions and $K_{\rm e4}$ decays
completed in 1979 [\cite{nag:etal}].  The values reported in that work are
\begin{equation}
    a_0^0 = 0.26 \pm 0.05\,\mu^{-1} \qquad {\rm and} \qquad 
    a_0^2 = -0.028 \pm 0.012\,\mu^{-1} \enspace . \label{eq:nagels}
\end{equation}
This result is clearly not precise enough to resolve the above theoretical
controversy.  We proceed to examine the more recent experiments and related 
attempts at extraction of new, more precise values of $a(\pi\pi)$.

\section{Experiments on Threshold $\pi$-$\pi$ Scattering}

As free pion targets cannot be fabricated, experimental evaluation of
$\pi\pi$ scattering observables is restricted to the study of a dipion
system in a final state of more complicated reactions.  Scattering lengths
are especially hard to determine since they require measurements close to
the $\pi\pi$ threshold, where the available phase space strongly reduces
measurement rates.  Over time several reactions have been studied or
proposed as a means to obtain near-threshold $\pi\pi$ phase shifts, such as
$\pi N\to\pi\pi N$, $K_{e4}$ decays, $\pi^+\pi^-$ atoms (pionium),
$e^+e^-\to\pi\pi$, etc.  In practice, only the first two reactions have so
far proven useful in studying threshold $\pi\pi$ scattering, although there
are ambitious plans to study pionium in the near future.  The main
experimental methods and current results are discussed below.

\subsection{$K_{\rm e4}$ Decays \label{sec:ke4}}

By most measures, the $K^+\to\pi^+\pi^-e^+\nu$ decay (called $K_{\rm e4}$)
provides the most suitable tool for the study of threshold $\pi\pi$
interactions.  The interaction takes place between two real pions on the
mass shell, the only hadrons in the final state.  The dipion invariant mass
distribution in $K_{\rm e4}$ decay peaks close to the $\pi\pi$ threshold,
and only two states, $l_{\pi\pi}=I_{\pi\pi}=0$ and
$l_{\pi\pi}=I_{\pi\pi}=1$, contribute appreciably to the process.  These
factors, as well as the well understood $V-A$ weak lagrangian giving rise
to the decay, favor the $K_{\rm e4}$ process among all others in terms of
theoretical uncertainties.  Measurements are, however, impeded by the low
branching ratio of the decay, $3.9 \times 10^{-5}$.

\begin{figure}[htb]
\vbox{\epsfig{figure=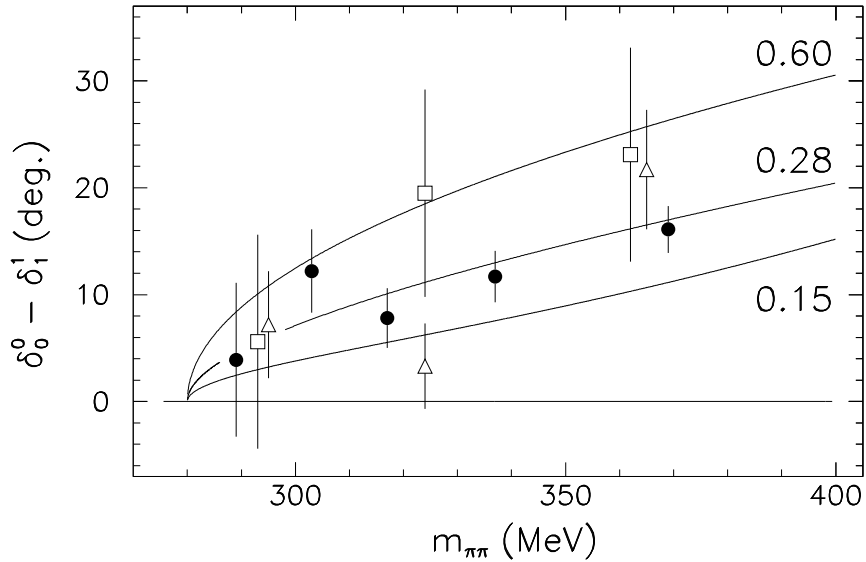,width=\textwidth} }
\caption{$\pi\pi$ phase shift difference $\delta^0_0 - \delta^1_1$
extracted from $K_{\rm e4}$ data is plotted against $m_{\pi\pi}$, the
dipion invariant mass.  Full circles: Rosselet et al.\
[\protect\cite{ros:etal}]; open squares Zylberstejn [\protect\cite{zylb}]; 
open triangles Beier et al.\ [\protect\cite{ros:etal}].  The three curves
correspond to phase shift solutions assuming three different values of
$a_0^0$, as noted.} 
\label{fig:ke4}
\end{figure}

Thus, $K_{\rm e4}$ decay data provide information on the $\pi$-$\pi$ phase
difference $\delta^0_0 - \delta^1_1$ near threshold.  The most recent
published $K_{\rm e4}$ experimental result was obtained by a Geneva--Saclay 
collaboration in the mid-1970's [\cite{ros:etal}].  Figure~\ref{fig:ke4}
summarizes the $\pi\pi$ phase shift information below 400 MeV derived from
all $K_{\rm e4}$ data published to date.  The curves in Fig.\,\ref{fig:ke4}
correspond to three different values of $a_0^0(\pi\pi)$, and illustrate the
relative insensitivity of the data to $a_0^0$ at the level of experimental
accuracy achieved by Rosselet et al.

Clearly, the available $K_{\rm e4}$ data are of insufficient accuracy.
Taken alone they provide a $\sim 35$\,\% constraint on $a_0^0$.  Only after
they are combined with $\pi\pi$ phase shifts extracted from peripheral $\pi
N\to\pi\pi N$ reactions (see Sect.\,\ref{sec:hi-en:pipin}) is it possible
to reduce the uncertainties to the level of about 20\,\%, as quoted in
(\ref{eq:nagels}).  However, new, substantially more precise $K_{\rm e4}$
data are expected in the near future (see Sect.\,\ref{sec:compare}).

We note that $K_{\rm e4}$ decays provide no information on the $I=2$
$\pi\pi$ phase shifts.  Hence, other reactions must be used to supplement
the $K_{\rm e4}$ data in order to study $I=2$ $\pi\pi$ scattering.

\subsection{Peripheral $\pi N\to\pi\pi N$ Reactions at High Momenta
\label{sec:hi-en:pipin}}

Goebel as well as Chew and Low showed in 1958/59 that particle production
in peripheral collisions can be used to extract information on the
scattering of two of the particles in the final state [\cite{che:low}].
This approach is, of course, useful primarily for the scattering of
unstable particles and has been used to great advantage in the study of the
$\pi\pi$ system.  Applied to the $\pi N\to\pi\pi N$ reaction, the
well-known Chew--Low formula,
\begin{equation}
  \sigma_{\pi\pi}(m_{\pi\pi}) = \lim_{t\to\mu^2}\ \left[ 
     {\partial^2\sigma_{\pi\pi N} \over \partial t\partial m_{\pi\pi} }
     \cdot {\pi \over \alpha f_\pi^2} \cdot {p^2(t-\mu^2)^2 \over
     tm_{\pi\pi}k} \right] \enspace , \label{eq:chew-low}
\end{equation}
relates $\sigma_{\pi\pi}(m_{\pi\pi})$, the cross section for pion-pion
scattering, to double differential $\pi N\to\pi\pi N$ cross section and
kinematical factors: $p$, momentum of the incident pion, $m_{\pi\pi}$, the
dipion invariant mass, $t$, the Mandelstam square of the 4-momentum
transfer to the nucleon, $k=(m_{\pi\pi}^2/4-\mu^2)^{1/2}$, momentum of the
secondary pion in the rest frame of the dipion, $f_\pi$, the pion decay
constant, and $\alpha=1$ or 2, a statistical factor involving the pion and
nucleon charge states.  The method relies on an accurate extrapolation of
the double differential cross section to the pion pole, $t=\mu^2$, in order
to isolate the one pion exchange (OPE) pole term contribution.  Since the
exchanged pion is off-shell in the physical region ($t<0$), this method
requires measurements under conditions which maximize the OPE contribution
and minimize all background contributions.  Thus, suitable measurements
require peripheral pion production at values of $t$ as close to zero as
possible, which becomes practical at incident momenta typically above
$\sim$3\,GeV/c.

\begin{figure}[htb]
\vbox{\epsfig{figure=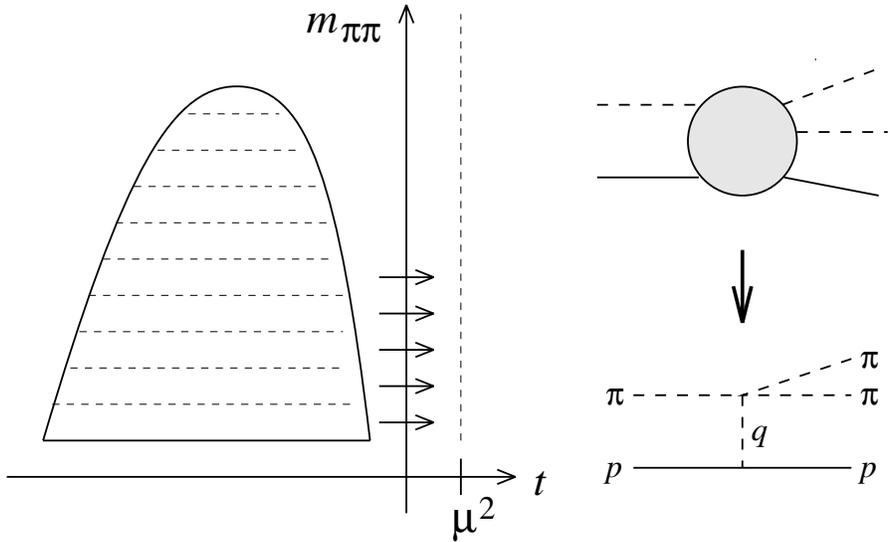,width=\textwidth} }
\caption{Illustration of the Chew--Low extrapolation in the $m_{\pi\pi}$
vs.\ $t$ plane (Chew--Low plane) to the pion pole $t=\mu^2$.  The
physical region of the data is bounded by the closed contour in the second
quadrant ($t<0$, $m_{\pi\pi}\geq 2\mu$).}
\label{fig:chew-low}
\end{figure}

The essential steps of the Chew--Low--Goebel procedure are illustrated in
Fig.\,\ref{fig:chew-low}.  The method relies on the assumption that the
dominant process in peripheral pion production (small $|t|$) is the OPE.
Since the pion has the smallest mass of all hadrons, the OPE pole lies
closest to the physical region ($t<0$) of any competing terms.  Thus, for
small $|t|$, the non-OPE background varies much more slowly than the OPE
term which, in turn, is proportional to \hbox{$t/(t-\mu^2)^2$}.  Hence,
measured  $\pi N\to\pi\pi N$ differential cross sections are plotted in the
so-called Chew--Low plane, $m_{\pi\pi}$ against $t$, as shown in
Fig.\,\ref{fig:chew-low}.  Data points are subdivided into bins (strips) of
$m_{\pi\pi}$ and for each bin the Chew--Low extrapolating function 
$F$, defined as
\begin{equation}
   F(s,t,m_{\pi\pi}) = {\partial^2 \sigma_{\pi\pi N}(s) \over \partial t\,
    \partial m_{\pi\pi}} \cdot {\pi \over f_\pi^2} \cdot
    {p^2(t-\mu^2)^2 \over t\, m_{\pi\pi}(m_{\pi\pi}^2 - 4\mu^2)^{1/2}}
    \enspace ,                                       \label{eq:f(s,t,m)}
\end{equation}
is extrapolated to the pion pole $t=\mu^2$ which lies outside of the
physical domain.  When angular momenta higher than $l=0$ contribute
significantly in the $\pi\pi$ system, $F(s,t,m_{\pi\pi})$ must first be
decomposed into spherical harmonics and the resulting amplitudes
extrapolated to the pion pole.

The Chew--Low method has been refined considerably over time, particularly
by Baton and coworkers [\cite{bat:etal}].  Crossing, Bose and isospin
symmetries, analyticity and unitarity, provide dispersion relation
constraints on the $\pi\pi$ phase shifts, the ``Roy equations''
[\cite{roy},\cite{bas:etal:1},\cite{bas:etal:2}].  Roy equations are
indispensable in evaluating $\pi\pi$ scattering lengths due to the
restricted phase space of peripheral $\pi N\to\pi\pi N$ reactions below
$m_{\pi\pi} \simeq 500$\,MeV; dispersion relations embodied in the Roy
equations make use of more accurate data available at higher $\pi\pi$
energies, compensating thus for the limitations of low-$m_{\pi\pi}$ data.

Since the Chew--Low--Goebel method relies on extrapolation in a
two-di\-men\-sio\-nal space, it requires kinematically complete data of
high quality, both in terms of measurement statistics and
resolution---these have been the limiting factors in all analyses to date.

The data base for these analyses has not changed essentially since the
early 1970's, and is dominated by two experiments, performed by the
Berkeley [\cite{pro:etal}] and CERN-Munich [\cite{gra:etal}] groups.  The
latter of the two measurements has much higher statistics (300 k events
compared to 32 k in the Berkeley experiment).  A comprehensive
analysis of this data base, with addition of the Geneva--Saclay $K_{\rm
e4}$ data, was performed by Nagels et al.\ [\cite{nag:etal}], as discussed
in Sect.~\ref{sec:ke4}.  The resulting values of $a_0^{0,2}$ are given in
(\ref{eq:nagels}). 

There have been other Chew--Low type analyses since 1979.  One, performed
by the Kurchatov Institute group in 1982, was based on a set of some 35,000
$\pi N\to\pi\pi N$ events recorded in bubble chambers [\cite{ale:etal}].
Patarakin, Tikhonov and Mukhin, members of the same group, recently updated
the 1982 analysis by including available data on the $\pi N\to\pi\pi\Delta$
reaction, as well as the published $K_{\rm e4}$ data [\cite{pat:etal}].
The resulting s-wave $\pi\pi$ scattering lengths were found to be bounded
by
\begin{equation}
   0.205\,\mu^{-1} < a_0^0 < 0.270\,\mu^{-1} \quad {\rm and} \quad
   -0.048\,\mu^{-1} < a_0^2 < -0.016\,\mu^{-1} \enspace . \label{eq:patarak}
\end{equation}
Although the above limits on $a_0^0$ carry slightly smaller uncertainties
than the generally accepted $a_0^0$ value of Nagels et al.\ listed in
(\ref{eq:nagels}), the result of Patarakin et al.\ still cannot
exclude one of the two competing pictures of chiral symmetry breaking
(strong vs.\ weak scalar quark condensate, as discussed in
Sect.\,\ref{sec:motiv}).  The central value, though, is lower than in
(\ref{eq:nagels}), more in line with the conventional, strong condensate 
picture that leads to the standard ChPT two-loop prediction of $a_0^0
\simeq 0.21\,\mu^{-1}$.

At this point it is worth to note a recent analysis by the Cracow group of
old unpublished CERN-Cracow-Munich $\pi^-\overrightarrow{p}\to\pi^-\pi^+n$
data at 17.2\,GeV, measured on a transversely polarized proton target
[\cite{kam:etal}].  The $m_{\pi\pi}$ range of this study is from 610 to
1590\,MeV.  In their analysis the Cracow group used a relativistic coupled
channel Lippmann-Schwinger treatment of the $\pi\pi$ and $\overline{K}K$
systems.  Results of the analysis of two data sets yielded values of
$a_0^0$ substantially lower than any discussed above:
\begin{equation}
      a_0^0 = \cases{0.172 \pm 0.008\,\mu^{-1} & for data set 1, \cr
                     0.174 \pm 0.008\,\mu^{-1} & for data set 2. \cr} 
                                                         \label{eq:kamin}
\end{equation}
This interesting analysis may have been affected adversely by the way the
original CERN-Cracow-Munich data were preserved.  Nevertheless, like the
work of the Kurchatov Institute group, this work points out that peripheral
$\pi N\to\pi\pi N$ data may indeed favor a lower value of $a_0^0$ than
indicated by the presently available $K_{\rm e4}$ data.

It is regrettable that new high energy ($E_\pi > 3$\,GeV) peripheral $\pi
N\to\pi\pi N$ measurements are not planned in the future.  Therefore
much attention during the last decade has been devoted to the study of the
$\pi N\to\pi\pi N$ reaction at lower energies, $p_\pi \leq 500$\,MeV.
These results are discussed next.

\subsection{Inclusive $\pi N\to\pi\pi N$ Reactions Near Threshold
\label{sec:lo-en:pipin}}
 
Weinberg showed early on [\cite{wein:1}] that the OPE graph dominates the
$\pi N\to\pi\pi N$ reaction at threshold.  Subsequently, Olsson and Turner
constructed a soft-pion lagrangian containing only the OPE and contact
terms at threshold [\cite{ols:tur}].  This enabled them to introduce a
simple parametrization of the relation between the $\pi\pi$ and $\pi
N\to\pi\pi N$ threshold amplitudes.  Although this work was superseded by
the emergence and general acceptance of QCD, it did provide the impetus for
a number of inclusive measurements of $\pi N\to\pi\pi N$ total cross
sections near threshold.  Results of these studies published before 1995
are reviewed in detail in Ref.\,[\cite{poc:1}].  That data base has
remained unchanged, apart from small additions that are discussed below.

As in peripheral pion production at high energies, there are 5 charge
channels accessible to measurement,
\begin{equation}
   \pi^-p\to\cases{\pi^-\pi^+n \cr 
                   \pi^0\pi^0n \cr 
                   \pi^-\pi^0p \cr} \qquad {\rm and} \qquad 
   \pi^+p\to\cases{\pi^+\pi^0p \cr
                   \pi^+\pi^+n \cr} \enspace . 
\end{equation}
Total cross sections of the five reactions are described by only four
independent isospin amplitudes $A_{2I,I_{\pi\pi}}$, namely, $A_{31},
A_{32}, A_{10}$ and $A_{11}$, where $I$ is the total ($\pi p$) isospin and 
$I_{\pi\pi}$ is the isospin of the dipion system.  Two of the four
amplitudes vanish at threshold due to Bose symmetry.  Thus, the amplitudes
are, in principle, overconstrained by data; this redundancy is welcome
given how difficult absolute measurements near threshold are.

\begin{figure}[htb]
\vbox{\epsfig{figure=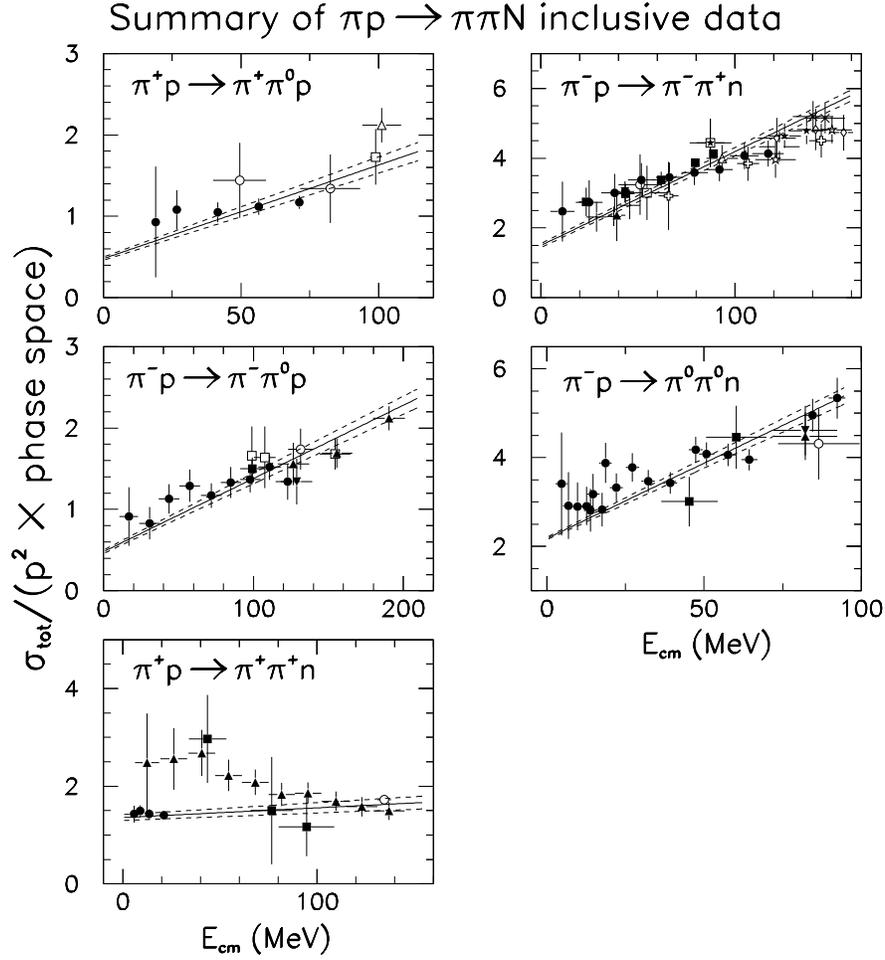,width=\textwidth} }
\caption{Summary of the published $\pi p\to\pi\pi N$ inclusive cross
section data, shown here with the factor ($p^2 \times$ phase space) divided
out, where $p$ is the beam momentum.  Solid lines shown in the figure are
the result of a simultaneous fit of the $\pi\pi N$ isospin amplitudes.  The
corresponding fit uncertainties are denoted by dashed lines.  For details
see Ref.\,[\protect\cite{poc:1}].}
\label{fig:incl_ampl}
\end{figure}

The amount and quality of available inclusive near-threshold $\pi
N\to\pi\pi N$ data, especially that collected since 1985, is impressive and
has resulted in relatively rigorous constraints on the $\pi\pi N$ isospin
amplitudes.  This is illustrated in Fig.\,\ref{fig:incl_ampl} which shows
the whole data base in the form of quasi-amplitudes obtained by removing
from the angle-integrated cross sections the uninteresting but strong
energy dependence due to the reaction phase space.

The current data base is increased compared to that of 1994 by the addition
of new, more precise $\pi^\pm p\to\pi^\pm\pi^+n$ cross sections very near
threshold from TRIUMF [\cite{lan:etal}].  The new measurements have
confirmed the same group's earlier published data [\cite{sev:etal}] on the
$\pi^+p\to\pi^+\pi^+n$ reaction, thus definitively invalidating older data
taken by the OMICRON collaboration at CERN [\cite{kern:etal}] (high-lying
points with large error bars in the bottom panel of
Fig.\,\ref{fig:incl_ampl}).

In spite of the relative abundance and high accuracy of the near-threshold
inclusive pion production data, their interpretation in terms of $\pi\pi$
scattering lengths has been plagued by theoretical uncertainties.  This
shortcoming has recently been successfully addressed within the framework
of the heavy baryon chiral perturbation theory (HBChPT) [\cite{ber:etal}].
Theoretical uncertainties limited the ability of this analysis to produce
a stringent constraint on the $I=0$ $\pi\pi$ channel.  However, the
HBChPT study did provide a restrictive new $I=2$ scattering length.  The two
results are:
\begin{equation}
    a_0^0 \simeq  0.21 \pm 0.07 \,\mu^{-1} \qquad {\rm and} \qquad
    a_0^2 = -0.031 \pm 0.007 \,\mu^{-1} \enspace .  \label{eq:hbchpt}
\end{equation}
The $a_0^0$ result was recently refined by Olsson who used the so-called
universal curve, a model-independent relation between $a_0^0$ and $a_0^2$
due to the forward dispersion relation or, equivalently, to the Roy
equations [\cite{olsson}].  Olsson found 
\begin{equation}
    a_0^0 = 0.235 \pm 0.03 \,\mu^{-1} \enspace .      \label{eq:olsson}
\end{equation}

Any analysis based on HBChPT cannot, however, be expected to result in
$\pi\pi$ scattering lengths different from the standard ChPT prediction
because the latter is built into the lagrangian used.

\subsection{Chew--Low analysis of Low Energy $\pi N\to\pi\pi N$ Data
\label{sec:lo-en:chewlow}} 

Given the theoretical uncertainties in the interpretation of inclusive $\pi
N\to\pi\pi N$ data near threshold, it was suggested some time ago to
apply the Chew--Low method to low energy $\pi N\to\pi\pi N$ data 
[\cite{poc:etal:1}].  Recently several exclusive $\pi N\to\pi\pi N$ data
sets suitable for such treatment have become available.  These are, in
the order in which they were measured: 

\indent (a) $\pi^-p\to\pi^0\pi^0n$ data from BNL [\cite{low:etal}], \\
\indent (b) $\pi^+p\to\pi^+\pi^0p$ data from LAMPF [\cite{poc:etal:2}], and
 \\  
\indent (c) $\pi^-p\to\pi^-\pi^+n$ data from TRIUMF [\cite{kerm:etal}]. 

We discuss below the current results of two analyses: first, of the LAMPF
E1179 data, set (b) above, by the University of Virginia group, and,
second, of the CHAOS data, set (c) above, by the TRIUMF group.

\subsubsection{Chew--Low Analysis of LAMPF E1179 Data.}  A
Virginia--Stanford--LAMPF team studied the $\pi^+p\to\pi^+\pi^0p$ reaction
at LAMPF at five energies from 190 to 260\,MeV [\cite{poc:etal:2}].  The
LAMPF $\pi^0$ spectrometer and an array of plastic scintillation telescopes
were used for $\pi^+$ and $p$ detection.  Three classes of exclusive events
were recorded simultaneously: $\pi^+\pi^0$ and $\pi^0p$ double
coincidences, and $\pi^+\pi^0p$ triple coincidences.  Since the acceptance 
of the apparatus and the backgrounds were significantly different for the
three classes of events, this experiment had a strong built-in consistency
check.  The $\pi^+p\to\pi^+\pi^0p$ reaction is sensitive only to the $I=2$
s-wave $\pi\pi$ scattering length.

\begin{figure}[htb]
\vbox{\hspace*{0.15\textwidth}
  \epsfig{figure=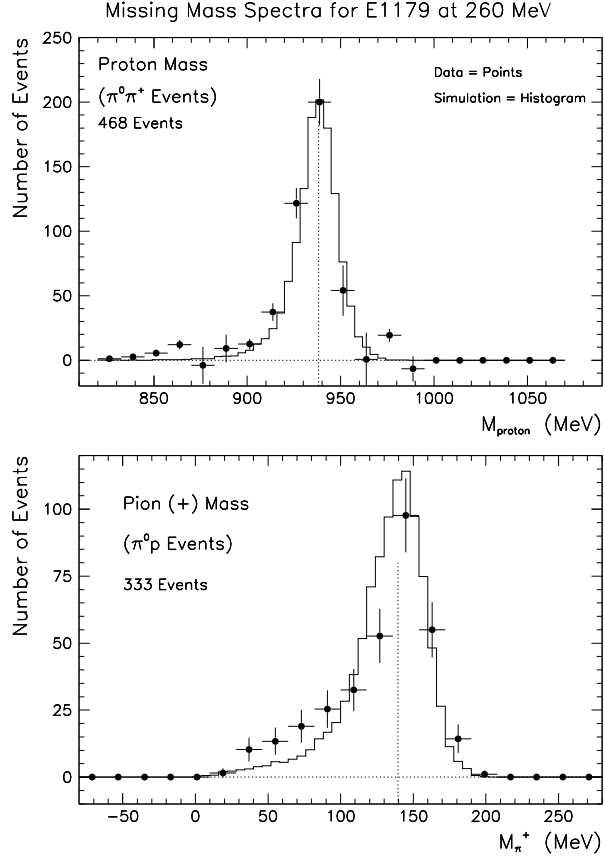,width=0.7\textwidth} }
\caption{Missing mass spectra for two classes of coincidence events,
$\pi^0\pi^+$ and $\pi^0p$, in the LAMPF E1179 $\pi^+p\to\pi^+\pi^0p$ data
set (full circles).  Histograms are the result of a detailed Monte Carlo
simulation of the apparatus and reaction.  Note the missing mass resolution
of $\sigma_p\simeq 11$\,MeV and $\sigma_\pi\simeq 17$\,MeV, respectively.
\label{fig:miss_mass:e1179} }
\end{figure}

Figure \ref{fig:miss_mass:e1179} illustrates the main source of difficulty
in this analysis, namely, the relatively broad energy resolution that
considerably smears the cross section data bins in a Chew--Low plot of
$m_{\pi\pi}$ against $t$.  Consequently, in order to obtain a physically
interpretable array of double differential cross section bins, a
complicated deconvolution procedure had to be implemented first
[\cite{bru}].  Limited counting statistics presented an additional
difficulty in the analysis, as it increased the uncertainties in both the
deconvolution procedure and in the final Chew--Low extrapolation.

\begin{figure}[htb]
\vbox{\epsfig{figure=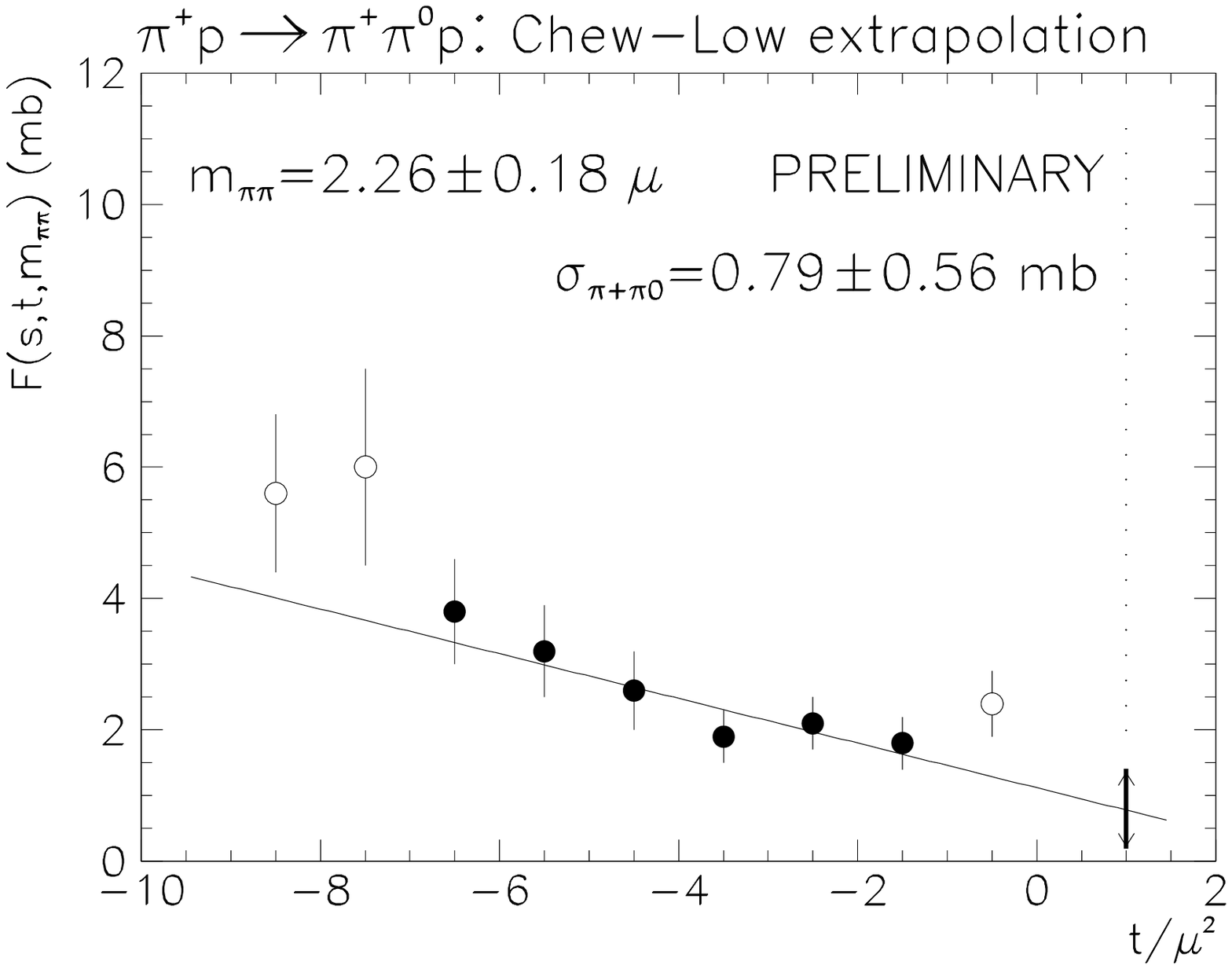,width=\textwidth} }
\caption{Chew-Low function $F(s,t,m_{\pi\pi})$ con\-struc\-ted from $\pi^+p
\to \pi^+\pi^0p$ exclusive cross sections at 260 MeV is plotted as a
function of $t$ along with a linear fit (preliminary).  Full circles: data
points included in the fit.  Open circles: data points excluded from the
fit. The ex\-tra\-po\-la\-ted value of the $\pi\pi$ total cross section at
$m_{\pi\pi} = 2.26 \pm 0.18\ \mu$ is indicated.}
\label{fig:chew-low:e1179}
\end{figure}

Preliminary results of this analysis for one bin of $m_{\pi\pi} = 2.26 \pm
0.18\,\mu$ are shown in Fig.\,\ref{fig:chew-low:e1179}.  Open circles in
the figure indicate data points excluded from the Chew--Low extrapolation
procedure due to large value of $|t|>6\,\mu^2$, where OPE is weak, and the
smallest $|t|$ point which has a large normalization uncertainty due to the
cross section deconvolution procedure.  The resulting $\pi\pi$ cross
section is $0.79 \pm 0.56$\,mb.  A proper procedure to extract
$a_0^2(\pi\pi)$ would be to include the new data point in a comprehensive
dispersion-relation $\pi\pi$ phase shift analysis.  Uncertainties in the
current analysis do not justify such an undertaking at this time.  However,
the precision one might expect from this result is illustrated by
evaluating $a_0^2$ from the above cross section datum directly.  Doing so
one obtains
\begin{equation}
    a_0^2 = -0.055 \pm 0.021\ \mu^{-1} \enspace ,
\end{equation}
which shows that the current status of this analysis does not provide a 
strong new constraint of the $\pi\pi$ phase shifts.  In comparison, the BNL
$\pi^-p\to\pi^0\pi^0n$ data, while having much higher event statistics, are
characterized by an even broader energy resolution and poorer coverage of
the low $|t|$ region critical for the Chew--Low extrapolation.

\subsubsection{Chew--Low Analysis of the CHAOS $\pi^-p\to\pi^+\pi^-n$
Data.}  The most significant development in this field in the past few
years has been the construction and operation of the Canadian High
Acceptance Orbit Spectrometer (CHAOS), a sophisticated new detector at
TRIUMF [\cite{chaos}].  This impressive device, composed of a number of
concentric cylindrical wire chamber tracking detectors and total energy
counters mounted between the poles of a large bending magnet, provides
nearly 360$^\circ$ of angular coverage for in-plane events, with excellent
acceptance for multi-particle events.  It is no surprise that the CHAOS
collaboration has very quickly measured the most comprehensive set of
exclusive in-plane $\pi^-p\to\pi^+\pi^-n$ cross sections below 300 MeV. 

\begin{figure}[htb]
\vbox{\epsfig{figure=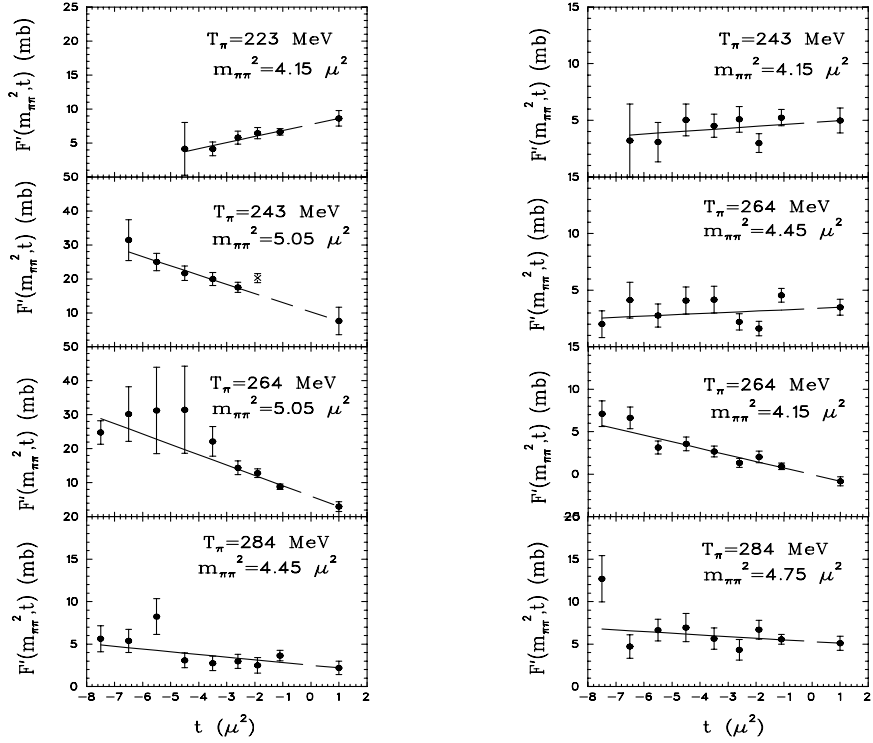,width=\textwidth} }
\caption{Plots of the Chew--Low extrapolation function $F(s,m_{\pi\pi},t)$
produced by the CHAOS group [\protect\cite{kerm:etal}].  The points at
$t=+\mu$ are deduced from extrapolation and yield the $\pi\pi$ cross
section.  Solid circles: data points used in the linear fit; crosses: data
points not used in the fit.  
\label{fig:chew-low:chaos} }
\end{figure}

The CHAOS $\pi^-p\to\pi^+\pi^-n$ data set covers four incident beam
energies between 223 and 284\,MeV.  Unlike the LAMPF and BNL measurements,
these data have an excellent energy resolution of $\sigma\simeq 4.8$\,MeV.
In order to carry out a Chew--Low analysis, the CHAOS collaborators binned
their data into an acceptance-corrected 10$\times$10$\times$10 lattice of
$m_{\pi\pi}^2$, $t$ and $\cos\theta$.  The $\cos\theta$ dimension was
integrated out, resulting in double-differential cross sections
d$^2\sigma$/d$m_{\pi\pi}^2$d$t$, which were used to construct the Chew--Low
extrapolating function $F(s,m_{\pi\pi},t)$, as given in
(\ref{eq:f(s,t,m)}).  A linear fit over a carefully selected interval in 
$t$ was made for every bin of $m_{\pi\pi}^2$.  The resulting fits and
linear extrapolation are shown in Fig.\,\ref{fig:chew-low:chaos}.

From the extrapolated values of $F(s,m_{\pi\pi},t)$ the authors extracted
$\pi\pi$ cross sections at six $\pi\pi$ energies in the range $m_{\pi\pi}^2
= 4.15$--5.65\,$\mu^2$ with uncertainties ranging from about 16\,\% at the
lowest energy to 63\,\% at the highest.  These $\pi\pi$ cross section data
were then added to the data base of Ref.\,[\cite{pat:etal}], and a Roy
equation constrained phase shift analysis was performed following the same
procedure as in Ref.\,[\cite{pat:etal}].  One parameter, $a_0^0$, was left
free to vary in the analysis.  Minimizing the $\chi^2$ of the fit, the
authors obtained
\begin{equation}
     a_0^0 = 0.206 \pm 0.013 \, \mu^{-1} \enspace , \label{eq:chaos_res}
\end{equation}
which would strongly confirm the validity of the standard ChPT and the
strong scalar quark condensate implied therein, at the same time ruling out
the possibility of the weak scalar quark condensate proposed by the Orsay
group [\cite{ste:etal:2},\cite{ste:etal:1}].

\subsubsection{Problems with the Chew--Low--Goebel Method at Low Energies?}
Bo\-lokhov et al.\ of the Sankt Petersburg State University have recently
performed a detailed study of the reliability of the Chew-Low method at low
energies using sets of synthetic $\pi N\to\pi\pi N$ ``data'' between 300
and 500 MeV/c [\cite{bol:etal}].  In this work the authors constructed data
sets with: (a) the OPE contribution only, (b) OPE + other allowed
mechanisms, (c) all mechanisms without the OPE.  Both linear and quadratic
Chew-Low extrapolation were used.  The authors found 25--35\,\% deviations
in the reconstructed OPE strength in case (a), 100--300\,\% deviations
under (b), and large ``OPE amplitude'' without any pion pole in the
synthetic data under (c).  This led the authors to conclude that ``\ldots
noncritical application [of the Chew--Low method] results in 100\,\%
theoretical errors, the extracted values being in fact random numbers
\ldots''

The quoted study is the first one to date to address theoretically the
validity of the Chew--Low--Goebel method in the low energy regime where
this technique has not been traditionally applied.  Given the complex
nature of the issue, it would be premature to write off using the method at
low energies altogether.  Clearly, a critical examination of the problem is
strongly called for.  However, before the matter is finally resolved, we
cannot accept the CHAOS result in (\ref{eq:chaos_res}) as definitive, in
spite of the high precision of the new CHAOS data, and of the elegance of
the analysis.

Further grounds for caution regarding the Chew--Low--Goebel method at low
energies are found in the pronounced pion beam energy dependence of the
extrapolated $\pi\pi$ cross sections in the lowest $m_{\pi\pi}^2$ bin, a
possible indication of a residual non-OPE background not properly removed
by the analysis.  It must be pointed out, however, that the authors found
that their result in (\ref{eq:chaos_res}) did not change significantly
when the lowest energy $\pi\pi$ cross section was dropped from the
analysis.  On the other hand, the same group's $\pi^+p\to\pi^+\pi^+n$ data
were incompatible with linear fits in terms of $F(s,m_{\pi\pi},t)$,
indicating a strong dominance of non-OPE processes in that reaction
channel.

\section{Summary of Current Results and Future Prospects
\label{sec:compare}} 

\begin{figure}[htb] 
\vbox{\epsfig{figure=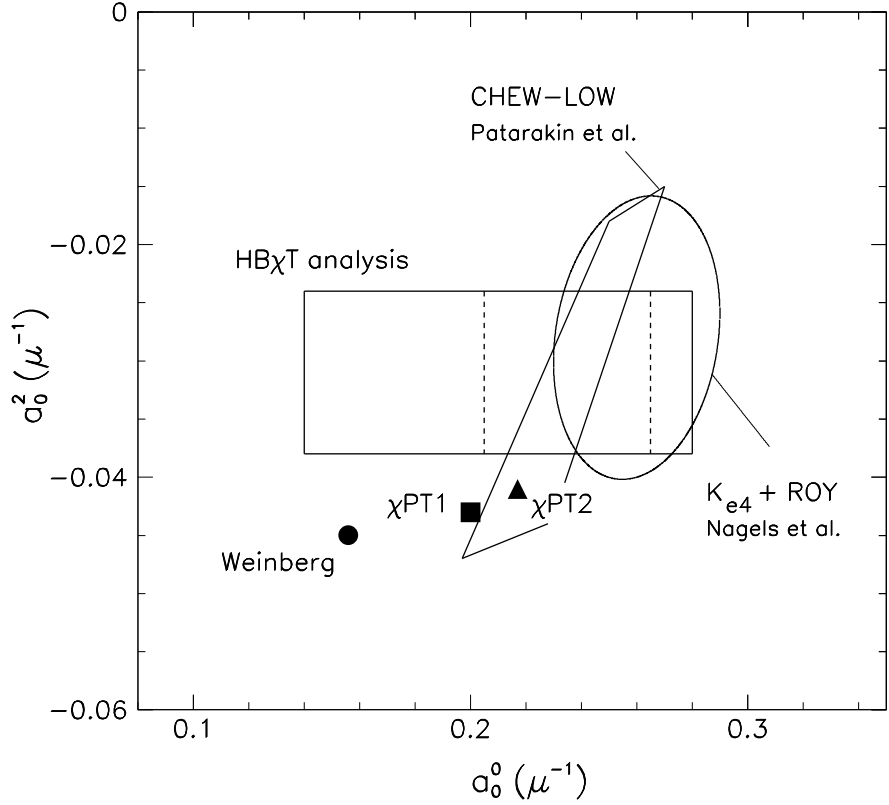,width=\textwidth} } 
\caption{Summary of $\pi\pi$ scattering length predictions: Weinberg's
tree-level result [\protect\cite{wein:1}] (full circle), ChPT one-loop
calculation [\protect\cite{gas:leu:2}] (full square), ChPT two-loop
calculation [\protect\cite{bij:col:1}] (full triangle), and analyses of
experimental data: Nagels et al. [\protect\cite{nag:etal}] (oval contour),
Patarakin et al. [\protect\cite{pat:etal}] (oblique quadrangular contour),
HBChPT analysis of Bernard et al. [\protect\cite{ber:etal}] (solid
rectangle), and Olsson's dispersion-relation constraint of the HBChPT
result [\protect\cite{olsson}] (dashed lines).  
\label{fig:univ_plot} } 
\end{figure}

Theoretical predictions and experimental results on the $\pi\pi$ scattering
lengths published to date are plotted in Fig.\,\ref{fig:univ_plot} in the
$a_0^2$ against $a_0^0$ plane.  

We note that the current analyses of the available $K_{\rm e4}$ and $\pi
N\to\pi\pi N$ data (excluding the not yet fully established low energy
application of the Chew--Low method) are not sufficiently accurate to
distinguish between the two scenarios of chiral symmetry breaking, i.e.,
between the standard strong scalar quark condensate picture and the one
with a weak $\langle 0|\bar{q}q|0\rangle$.

At the same time the available analyses seem to favor slightly higher
values of both $a_0^0$ and $a_0^2$ than the values predicted by standard
ChPT.

The threshold $\pi$-$\pi$ scattering experimental data base will improve
significantly in the near future as several new experiments, listed below,
bear fruit.  The same experiments are discussed in more detail elsewhere in
these Proceedings.
\smallskip

\noindent{\bf \boldmath $K_{\rm e4}$ Data from BNL E865.}  Recently
completed measurements carried out by the E865 collaboration at BNL have
resulted in more than $3\times10^5$ $K_{\rm e4}$ decay events on tape
[\cite{lowe}].  Since the analysis of these data had not progressed far at
the time of this writing, the final event statistics after the appropriate
cuts are applied remains to be determined.  For comparison, the data base
of Rosselet et al.\ consisted of 30,000 events, so a significant
improvement is expected from the BNL E865 work.
\smallskip

\noindent{\bf \boldmath $K_{\rm e4}$ Data from DA$\Phi$NE.}  The KLOE
detector at the Frascati $\phi$ factory DA$\Phi$NE will be used in an
ambitious program of measurement of the $K_{\rm e4}$ decay.  The expected
accuracy of the $\pi\pi$ phase shift difference $\delta_0^0 - \delta_1^1$
to be extracted from this work is 5\,\% [\cite{bail:fra}], i.e., almost an
order of magnitude improvement over the current result.
\smallskip

\noindent{\bf \boldmath Lifetime of the $\pi^+\pi^-$ Atom (CERN).}  The
DIRAC experiment at the SPS at CERN [\cite{dirac:cern}] grew out of the
first observation of the $\pi^+\pi^-$ atom (pionium) at the Serpuhov
laboratory [\cite{pionium}].  The DIRAC project relies on the Lorentz
boost of relativistic pionium to measure the lifetime of the pionium atom
to 10\,\% accuracy.  This, in turn, will constrain the $\pi\pi$ scattering
length difference $|a_0^0-a_0^2|$ with 5\,\% accuracy.  In this respect,
the pionium and $K_{\rm e4}$ decay experiments are complementary, as the
latter provide no direct information on $a_0^2$.
\smallskip

As has been noted, further theoretical work is required to make use of the
existing $\pi N\to\pi\pi N$ data, in particular to clarify the
applicability of the Chew--Low--Goebel method at low energies.
Additionally, better understanding of the electromagnetic corrections will
be necessary in order to take full advantage of the forthcoming $K_{\rm
e4}$ and pionium data.  Thus, the next few years will be interesting on
both the experimental and theoretical fronts.
\medskip

The author wishes to thank A. A. Bolokhov, E. Frle\v{z}, O. O. Patarakin,
M. E. Sevior and G. R. Smith for substantive discussions and for graciously
providing access to results of their ongoing work.  This work has been
supported by a grant from the U.S. National Science Foundation.

% 
% ---- Bibliography ---- 
% 


\begin{thebibliography}

\bibitem{}{wein:1}{1} [1] S. Weinberg, Phys.\,Rev.\,Lett. {\bf 17}, 616
(1966); {\it ibid.} {\bf 18}, 188 (1967).

\bibitem{}{gas:leu:1}{2} [2] J. Gasser and H. Leutwyler, Ann.\,Phys. (N.Y.)
{\bf 158}, 142 (1984); Nucl.\,Phys. {\bf B250}, 465 (1985).

\bibitem{}{gas:leu:2}{3} [3] J. Gasser and H. Leutwyler, Phys.\,Lett. {\bf
B125}, 325 (1983).

\bibitem{}{bij:col:1}{4} [4] J. Bijnens, et al.,
%G. Colangelo, G. Ecker, J. Gasser and M. E. Sainio, 
Phys.\,Lett. {\bf B374}, 210 (1996); Nucl.\,Phys. {\bf B508}, 263 (1997).

\bibitem{}{ste:etal:1}{5} [5] J. Stern, H. Sazdjian and N. H. Fuchs,
Phys.\,Rev.\,D {\bf 47}, 3814 (1993); M. Knecht, B. Moussallam and
J. Stern, Nucl.\,Phys. {\bf B429}, 125 (1994).

\bibitem{}{ste:etal:2}{6} [6] N. H. Fuchs, H. Sazdjian and J. Stern,
Phys.\,Lett. {\bf B238}, 380 (1990)

\bibitem{}{kne:mou:1}{7} [7] M. Knecht, et al.,
%B. Moussallam, J. Stern and N. H. Fuchs, 
Nucl.\,Phys. {\bf B457}, 513 (1995).

\bibitem{}{kne:mou:2}{8} [8] M. Knecht, et al.,
%B. Moussallam, J. Stern and N. H. Fuchs, 
Nucl.\,Phys. {\bf B471}, 445 (1996). 

\bibitem{}{nag:etal}{9} [9] M. M. Nagels, et al.,
% T. A. Rijken, J. J. De Swart, G. C. Oades, J. L. Petersen, A. C. Irving,
% C. Jarlskog, W. Pfeil, H. Pilkuhn and H. P. Jakob, 
Nucl.\,Phys. {\bf B147}, 189 (1979).

\bibitem{}{ros:etal}{10} [10] L. Rosselet, et al., Phys.\,Rev.\,D 
{\bf 15}, 574 (1977).

\bibitem{}{zylb}{11} [11] A. Zylberstejn, Ph.D. thesis, University of
Paris, Orsay, 1972.

\bibitem{}{bei:etal}{12} [12] E. W. Beier et al.,
Phys.\,Rev.\,Lett. {\bf 29}, 511 (1972); {\it ibid.} {\bf 30}, 399 (1973).

\bibitem{}{che:low}{13} [13] C. J. Goebel, Phys.\,Rev.\,Lett. {\bf 1}, 337
(1958);  G. F. Chew and F. E. Low, Phys.\,Rev. {\bf 113}, 1640 (1959).

\bibitem{}{bat:etal}{14} [14] J. B. Baton, G. Laurens and J. Reignier,
Phys.\,Lett. {\bf 33B}, 525 (1970).

\bibitem{}{roy}{15} [15] S. M. Roy, Phys.\,Lett. {\bf35B}, 353 (1971).

\bibitem{}{bas:etal:1}{16} [16] J. L. Basdevant, J. C. Le Guillou and
H. Navelet, Nuovo Cim. {\bf 7A}, 363 (1972).  

\bibitem{}{bas:etal:2}{17} [17] J. L. Basdevant, C. G. Froggatt and
J. L. Peterson, Nucl.\,Phys. {\bf B72}, 413 (1974). 

\bibitem{}{pro:etal}{18} [18] S. D. Protopopescu et al.,
Phys.\,Rev. D {\bf 7}, 1279.
(1973)

\bibitem{}{gra:etal}{19} [19] G. Grayer et al., Nucl.\,Phys. {\bf
B75}, 189 (1974).

\bibitem{}{ale:etal}{20} [20] E. A. Alekseeva et al.,
Zh. Eksp. Teor. Fiz. {\bf 82}, 1007 (1982) [Sov. Phys. JETP {\bf 55}, 591
(1982)].

\bibitem{}{pat:etal}{21} [21] O. O. Patarakin, V. N. Tikhonov and
K.N. Mukhin, Nucl.\,Phys. {\bf A598}, 335 (1996).

\bibitem{}{kam:etal}{22} [22] R. Kami\'nski, L. Le\'sniak and
J. P. Maillet, Phys.\,Rev.\ D {\bf 50}, 3145 (1994); R. Kami\'nski and 
L. Le\'sniak, Phys.\,Rev.\ C {\bf 51}, 2264 (1995).

\bibitem{}{ols:tur}{23} [23] M. G. Olsson and L. Turner,
Phys.\,Rev.\,Lett. {\bf 20}, 1127 (1968); Phys.\,Rev. {\bf 181} 2141 (1969).

\bibitem{}{poc:1}{24} [24] D. Po\v{c}ani\'c, in {\it ``Chiral Dynamics,
Theory and Experiment''}, A. M. Bernstein and B. R. Holstein, eds.,
Lect.\,Notes in Phys.\,Vol.\ 452, (Springer Verl., 1995) 95. 

\bibitem{}{lan:etal}{25} [25] J. B. Lange, et al., TRIUMF preprint
(1997).

\bibitem{}{sev:etal}{26} [26] M. E. Sevior, et al.,
Phys.\,Rev.\,Lett. {\bf 66}, 2569 (1991).

\bibitem{}{kern:etal}{27} [27] G. Kernel et al., Z.\,Phys. {\bf C48},
201 (1990).

\bibitem{}{ber:etal}{28} [28] V. Bernard, N. Kaiser and Ulf G. Meissner,
Nucl.\,Phys. {\bf B457}, 147 (1995).

\bibitem{}{olsson}{29} [29] M. G. Olsson, Phys.\,Lett. {\bf B410}, 311
(1997).

\bibitem{}{poc:etal:1}{30} [30] D. Po\v{c}ani\'c et al., proposal for
LAMPF experiment E1179 (1989).

\bibitem{}{low:etal}{31} [31] J. Lowe, et al., Phys.\,Rev.\,C {\bf 44}
956 (1991).

\bibitem{}{poc:etal:2}{32} [32] D. Po\v{c}ani\'c, et al., 
Phys.\,Rev.\,Lett. {\bf 72}, 1156 (1994); E. Frle\v{z}, Ph.\,D. Thesis,
Univ.\ of Virginia, 1993 (Los Alamos Report LA-12663-T, 1993).

\bibitem{}{kerm:etal}{33} [33] M. Kermani, et al., TRIUMF preprint
(1997).

\bibitem{}{bru}{34} [34] S. E. Bruch, M.Sc.\,Thesis, Univ.\ of Virginia
(1995). 

\bibitem{}{chaos}{35} [35] G. R. Smith, et al., Nucl.\,Instrum.\,Meth. {\bf
A362}, 349 (1995). 

\bibitem{}{bol:etal}{36} [36] A. A. Bolokhov, M. V. Polyakov and
S. G. Sherman, e-print hep-ph/9707406 (1997).

\bibitem{}{lowe}{37} [37] J. Lowe, contribution to this Proceedings
(1997).

\bibitem{}{bail:fra}{38} [38] M. Baillargeon and P. J. Franzini,
``Accuracies of $K_{\ell 4}$ Parameters at DA$\Phi$NE'', in the Second
DA$\Phi$NE Handbook, L. Maiani, N. Paver and G. Pancheri, eds., 1995.

\bibitem{}{dirac:cern}{39} [39] B. Adeva et al., ``Lifetime measurement of
$\pi^+\pi^-$ atoms to test low energy QCD predictions'', proposal to the
SPSLC, CERN/SPSLC 95-1 (1995).

\bibitem{}{pionium}{40} [40] L. G. Afanasev, et al., Phys.\,Lett. {\bf
B308}, 200 (1993).
 
\end{thebibliography}
\end{document}